\documentclass[runningheads]{llncs}
\usepackage[T1]{fontenc}
\usepackage{graphicx}
\usepackage[backend=bibtex,style=numeric-comp,giveninits=true,doi=false,isbn=false,url=false,eprint=false,maxbibnames=99]{biblatex}
\usepackage[dvipsnames]{xcolor}
\usepackage{amsmath}
\usepackage{booktabs} 
\usepackage{siunitx}
\begin{document}
\title{Large deformation diffeomorphic \\cardiac strain mapping}
\titlerunning{Large deformation diffeomorphic cardiac strain mapping}
%
\author{Beatrice Moscoloni \inst{1,2}\and
Patrick Segers \inst{2} \and
Mathias Peirlinck\inst{1}}

\authorrunning{B. Moscoloni et al.}

\institute{Delft University of Technology, the Netherlands \and Ghent University, Belgium}

\maketitle              
\begin{abstract}
Cardiac deformation is a crucial biomarker for the evaluation of cardiac function. Current methods for estimating cardiac strain might underestimate local deformation due to through-plane motion and segmental averaging. Mesh-based mapping methods are gaining interest for localized analysis of cardiac motion and strain, yet they often do not consider important properties of cardiac tissue. In this work, we propose an extension of the large deformation diffeomorphic metric mapping framework to incorporate near incompressibility into the loss function that guides the mapping. As such, our mechanically regularized mLDDMM allows for accurate and mechanically coherent estimation of volume displacement and strain tensors from time-resolved three-dimensional meshes. We benchmark our method against the results of a finite element simulation of cardiac contraction and find a very good agreement between our estimation and the simulated ground truth. Our method forms a promising technique to extract volume displacement and strain tensors from time-resolved meshes while accounting for the incompressibility of cardiac tissue. 

\end{abstract}
\section{Introduction}
Cardiac strain analysis plays a critical role in diagnosing pathological heart conditions, with early measurements serving as crucial biomarkers for cardiac dysfunction \cite{Lu2020}. 
Traditional methods, such as speckle-tracking echocardiography and cardiac magnetic resonance imaging feature tracking, are widely used but suffer from limitations. 
These include segmental averaging, which may obscure localized deformations, manual segmentation, and through-plane motion effects that can introduce variability in strain estimation \cite{Nabeshima2020, Xu2022,Pentenga2023,Smiseth2024}. 
Additionally, these techniques often rely on vendor-specific software, which limits broader applicability  \cite{Mukherjee2024}.
Mesh-based motion and strain analysis offers a more precise alternative for three-dimensional motion tracking of cardiovascular structures \cite{Biffi2017,Peirlinck2019BC,Kolawole2023,Meng2024,Mukherjee2024,Folco2025}. 
Recent advances in computer vision enable the automated extraction of detailed three-dimensional meshes from medical images \cite{Moscoloni2025}, although challenges remain to maintain a consistent mesh structure and vertex correspondence during displacement field estimation \cite{Biffi2017,Kong2021, Meng2024}. 
Large deformation diffeomorphic metric mapping (LDDMM) has been explored for motion tracking of cardiac structures, but it primarily focuses on surface meshes and does not capture the expected volumetric deformation of myocardial tissue \cite{Biffi2017, Berberoglu2019,Moscoloni2025}.
In this work, we propose a novel method for extracting volumetric displacement and strain tensors from time-resolved meshes over a cardiac cycle. 
Our enriched LDDMM framework, which we call \textit{mLDDMM}, incorporates a near-incompressibility constraint into the loss function that drives the mapping, ensuring accurate and mechanically coherent mapping - and reliable estimates - of volumetric deformation and strain tensors. 
We validate our approach with ground-truth data from a finite element simulation of cardiac contraction.
\section{Methods}
\subsection{Synthetic data generation: finite element analysis of cardiac mechanics}
We set up a high-fidelity, four-chamber human heart model with proximal vasculature to simulate realistic cardiac mechanics throughout the cardiac cycle. 
The model, derived from high-resolution magnetic resonance and computed tomography imaging of a healthy, 50th percentile U.S. male \cite{Baillargeon2014, Peirlinck2021, Zygote2014}, captures detailed anatomical features, including chambers, valves, chordae tendinae, and papillary muscles. 
We assign the heart muscle's intrinsic myofiber architecture using rule-based algorithms, informed by histological observations and diffusion tensor magnetic resonance images \cite{Bayer2012, Lombaert2012}. 
Specifically, we implement the Laplace-Dirichlet rule-based algorithm  and impose the myofibers' helical angle to vary transmurally from \ang{-41} in the outer wall to \ang{+66} in the inner wall \cite{Peirlinck2019BC}. 
We leverage a reconstructed geometry at 70\% diastole where we discretize the myocardial wall using 303,870 quadratic tetrahedral elements and 69,910 nodes. 
Since the geometry of the heart is constructed while it is already hemodynamically loaded, we computationally account for the in vivo prestress at the beginning of the simulation using an inverse elastostatics method \cite{Peirlinck2018}.

\smallskip
\noindent To model cardiac tissue mechanics, we enforce mass and momentum conservation laws and incorporate Holzapfel and Ogden's orthotropic invariant-based constitutive model for passive elasticity \cite{Holzapfel2009,Martonova2024,Peirlinck2024c} and a time-varying elastance model for active contraction \cite{Walker2005, Peirlinck2019BC}. 
We impose homogeneous Dirichlet boundary conditions at vascular outlets \cite{Peirlinck2019BC} and prescribe physiological pre- and after-loading conditions through a closed-loop lumped-parameter network model \cite{Baillargeon2014}. 
We sequentially calibrate passive and active material properties based on average in vivo pressure volume data for healthy subjects \cite{Peirlinck2021}.
Specifically, we separately scale the combined sets of linear and exponential constitutive parameters, identifying optimal scaling factors by minimizing the difference between the simulated and experimentally derived diastolic pressure-volume relationships constructed using the Klotz curve \cite{Peirlinck2019HF}. Next, we optimize maximum contractility to obtain healthy peak systolic interventricular pressures and stroke volumes.
Model validation is performed using clinical metrics such as ejection fraction and left ventricular twist \cite{Peirlinck2019BC,Peirlinck2021}.
Using the calibrated parameters, we simulate five cardiac cycles, after which the heart reaches cyclic steady-state equilibrium. 
Finally, we extract the ventricular displacement fields $\mathbf{u}$, which serve as ground-truth input data for our subsequent analyses. 
To ensure that the output of our deformation estimation and the ground truth deformation profile of our fifth steady-state cardiac cycle are comparable, we adjust the simulation output by subtracting the initial pre-stretch displacement from the entire displacement field. As such, both the estimation and the ground-truth consistently begin from the same zero-displacement baseline. 
\subsection{Volume displacement and deformation estimation}
Using the open-source LDDMM implementation in \textit{Deformetrica}, we map the time-resolved mesh that geometrically discretizes the reference myocardial ventricular solid domain  $\Omega_0$ to a subsequent target mesh of the displaced and deformed domain $\Omega$. 
The mapping process involves iteratively deforming an ambient space of equally spaced control points embedding the template, while optimizing a loss function $\mathcal{L}$ \cite{Bone2018, Glaunes2008, Moscoloni2025} which - \textit{classically} -  aims to minimize the Varifold distance $\mathcal{D}$ between the reference surface mesh  $\Gamma_0 = \partial \Omega_0$ and the target surface mesh $\Gamma = \partial \Omega$. 
The deformed collection of control points translates into a volumetric displacement field $\mathbf{u}(\mathbf{X})=x$ which maps material points $\mathbf{X} \in \Omega_0$ to their spatial positions $\mathbf{x} \in \Omega$ in the displaced and deformed configuration.
\smallskip

\noindent Given that the optimization of the surface-distance-based loss function does not intrinsically obey mechanical constraints of deformation within the tissue, we \textit{expand} the classic LDDMM formulation to a \textit{mechanically regularized mLDDMM} framework by including \textit{mechanical constraints} in the loss function.
More specifically, we constrain the solution space to only produce displacement fields $\mathbf{u}$ that abide by the natural near-incompressible behavior of myocardial tissue.
Towards this goal, during each mLDDMM-iteration $i$ we leverage the proposed Varifold distance-based surface-matching $\mathbf{u}_{\Gamma,i}$ to solve the boundary value problem for Laplace's equation 
\begin{equation}
\begin{aligned}
&\nabla^2 \mathbf{u}_i=0 \quad  &\text{in } \, &\Omega \\
&\mathbf{u}=\mathbf{u}_{\Gamma ,i}\quad &\text{on } &\partial\Omega = \Gamma.
\label{eqn:volume_displacement}
\end{aligned}
\end{equation}
to compute the volumetric deformation gradient for the whole domain as:
\begin{equation}
\mathbf{F}_i = \nabla{\mathbf{u}_i} + \mathbf{I}.
\label{eqn:deformation_field}
\end{equation}
Next, we compute the mechanical incompressibility regularization term $\mathcal{I}_i$ which penalizes deviations of the determinant of the deformation gradient $\mathbf{F}$ from 1, weighted by the element volume in a finite element mesh:
\begin{equation}
\mathcal{I}_i = \sum_e V_e \left( \det\left(\mathbf{F}_{i,e}\right) - 1\right)^2 
\end{equation}
where \( V_e \) is the element volume, \( \mathbf{F}_{i,e} \) is the deformation gradient of iteration $i$, and \( \det\left(\mathbf{F}_{i,e}\right) \) measures local volume change.
Our resulting \textit{mechanically regularized loss} $\mathcal{L}$ reads:
\begin{equation}
\mathcal{L} = \mathcal{D} + \alpha \cdot \mathcal{I}
\label{eqn:loss_function_enhanced}
\end{equation}
where the hyperparameter $\alpha$ controls the relative strength of the mechanical regularization with respect to the Varifold distance $\mathcal{D}$ during optimization.
We adopt an adaptive weighting strategy for $\alpha$, which strongly penalizes proposed $\mathbf{u}_i$ solutions that surpass a threshold incompressibility constraint $\mathcal{I}_t$:
\begin{equation}
    \alpha = \alpha_- + (\alpha_+ - \alpha_-) \cdot \sigma \left(\frac{\mathcal{I}_i-\mathcal{I}_t}{\mathcal{N}}\right)
\end{equation}
where we set $\alpha_-=0$, $\alpha_+=200$, $\mathcal{I}_t=1000$, and the regularization term $\mathcal{N}=20000$.
We apply our expanded mechanically regularized mLDDMM framework on the endo- and epicardial surface meshes deduced from our ground-truth cardiac mechanics simulation at end diastole and end systole, i.e. our reference and target configurations, respectively. 
Here, we use a grid of control points defined by the hyperparameter $\lambda_W = $10mm and set the stiffness-hyperparameter of allowed deformation $\lambda_V = $5mm \cite{Moscoloni2025}. We run this optimization until convergence, namely we stop the optimization when the relative decrease of the loss function during an iteration of the estimation is below $1e^{-4}$.

\subsection{Strain tensor computation}
Given the optimized $\mathbf{u} = \mathbf{u}_{\textrm{opt}}$ and the corresponding deformation gradient $\mathbf{F}$, we calculate the \textit{true} strain tensor in the current configuration as \cite{abaqus_user_manual}: 

\begin{equation}
\mathbf{\epsilon} = 
\ln\left(\sqrt{\mathbf{F}\cdot\mathbf{F}^{T}} \right) = 
\ln \left( \sqrt{\mathbf{V}\mathbf{R}\cdot\mathbf{R}^{T}\mathbf{V}^{T}} \right) = 
\ln \left( \sqrt{\mathbf{V}\cdot\mathbf{V}} \right)
\label{eqn:le-strain}
\end{equation}
We use the FE-derived displacement field to calculate a ground-truth deformation gradient as in Equation \ref{eqn:deformation_field}, and calculate the FE-derived strain measurements following Equation \ref{eqn:le-strain}.
\smallskip

\noindent In line with similar applications \cite{Mukherjee2024}, we validate the estimated strain tensor against FE-derived strain measurements using the mean squared error (MSE) metric. We calculate the MSE for each volumetric element as:
\begin{equation}
    MSE_i = ||\epsilon_i - \hat{\epsilon}_i||^2
\end{equation}
to obtain the field of the MSEs and evaluate the ability of our mesh-derived strain extraction method to recover regional cardiac deformation patterns. 
Additionally, we analyze the 99th percentile of the MSEs for each component over all elements to obtain a global error metric, while excluding the effect of outliers. 
\section{Results}
\subsection{Displacement estimation}
%
\begin{figure}[h]
    \centering
    \includegraphics[width=1.0\linewidth]{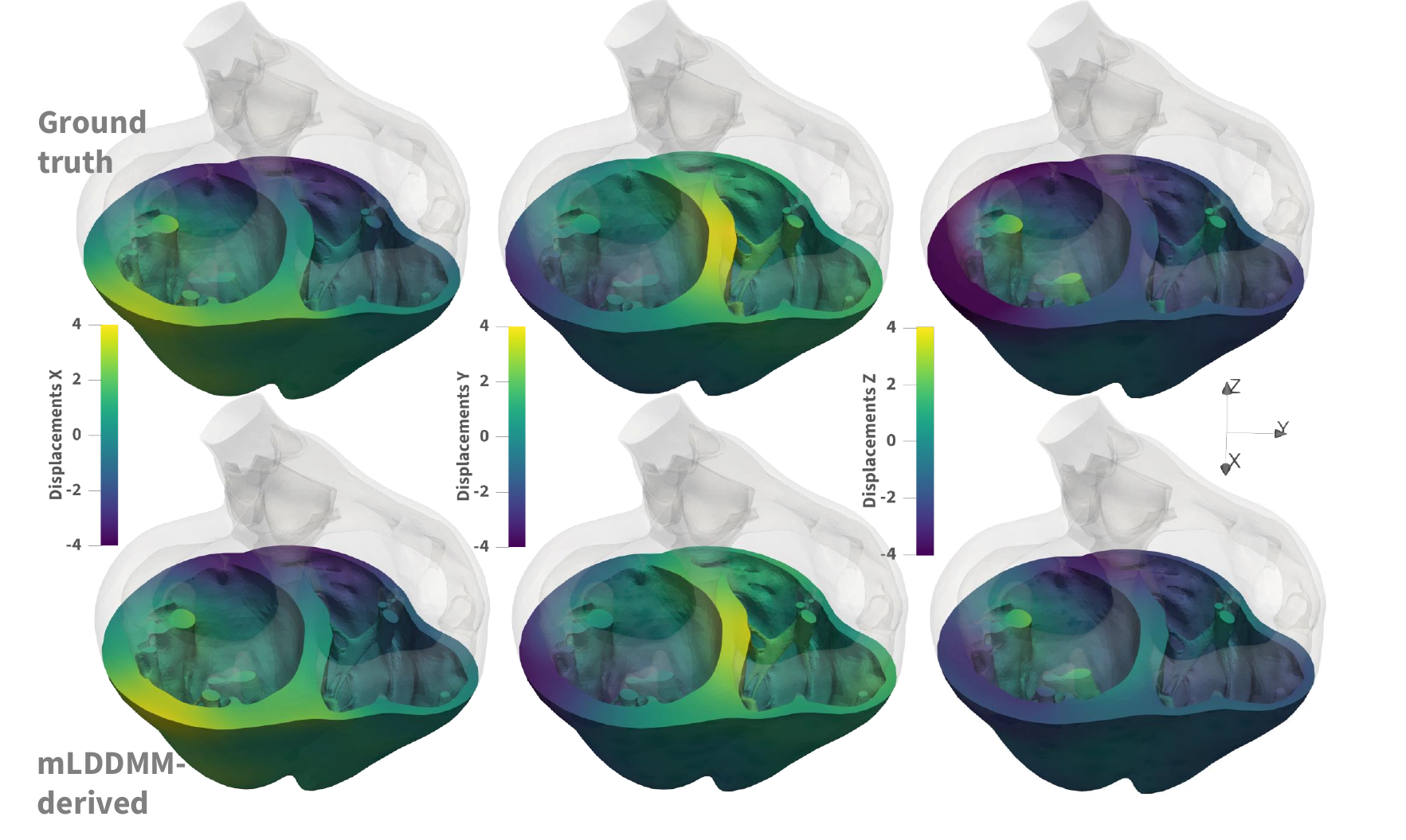}
    \caption{\textbf{mLDDMM displacement estimation power.} Ground truth FE-computed (top row) and mLDDMM-derived (bottom row) displacement fields. The first, second, and third column display deformation in the x-, y-, and z-directions respectively.}
    \label{fig:u_comparison}
\end{figure}
Figure \ref{fig:u_comparison} compares ground truth FE-computed and mLDDMM-estimated volumetric displacement fields. In the x- and y- directions, mLDDMM displacement estimates closely match the ground-truth finite element analysis solution.
mLDDMM slightly underestimates z-directional displacement in the left ventricular free wall and septum, albeit closely reproducing the overall displacement pattern. 

\begin{figure}[h]
    \centering
    \includegraphics[width=1.0\linewidth]{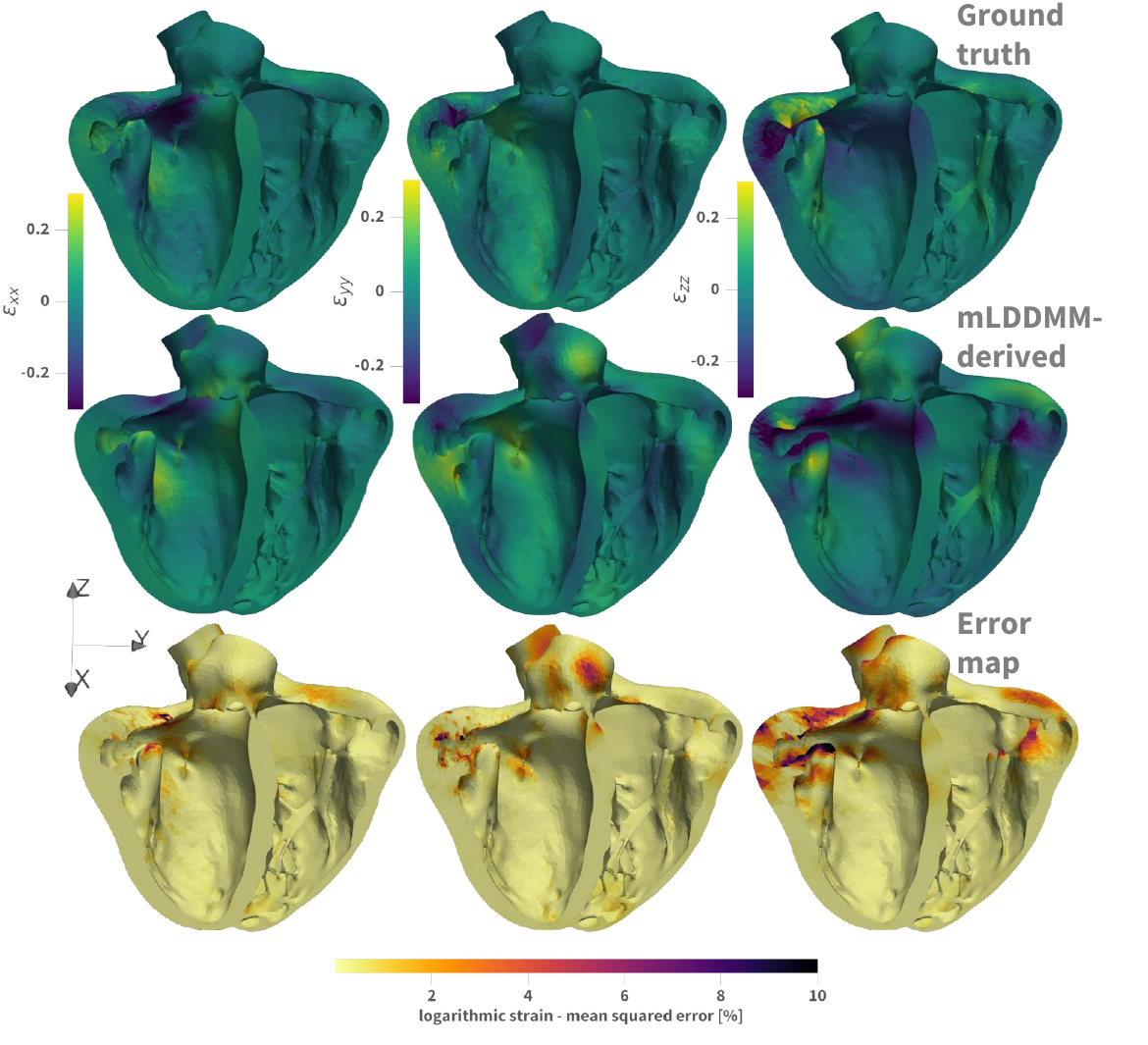}
    \caption{\textbf{mLDDMM strain estimation power, long-axis.} Ground truth FE-computed and mLDDMM-derived logarithmic strain comparison, long-axis view.}
    \label{fig:le_comparison_long_axis_errors}
\end{figure}
\subsection{Strain estimation}
\begin{figure}[h]
    \centering
    \includegraphics[width=1.0\linewidth]{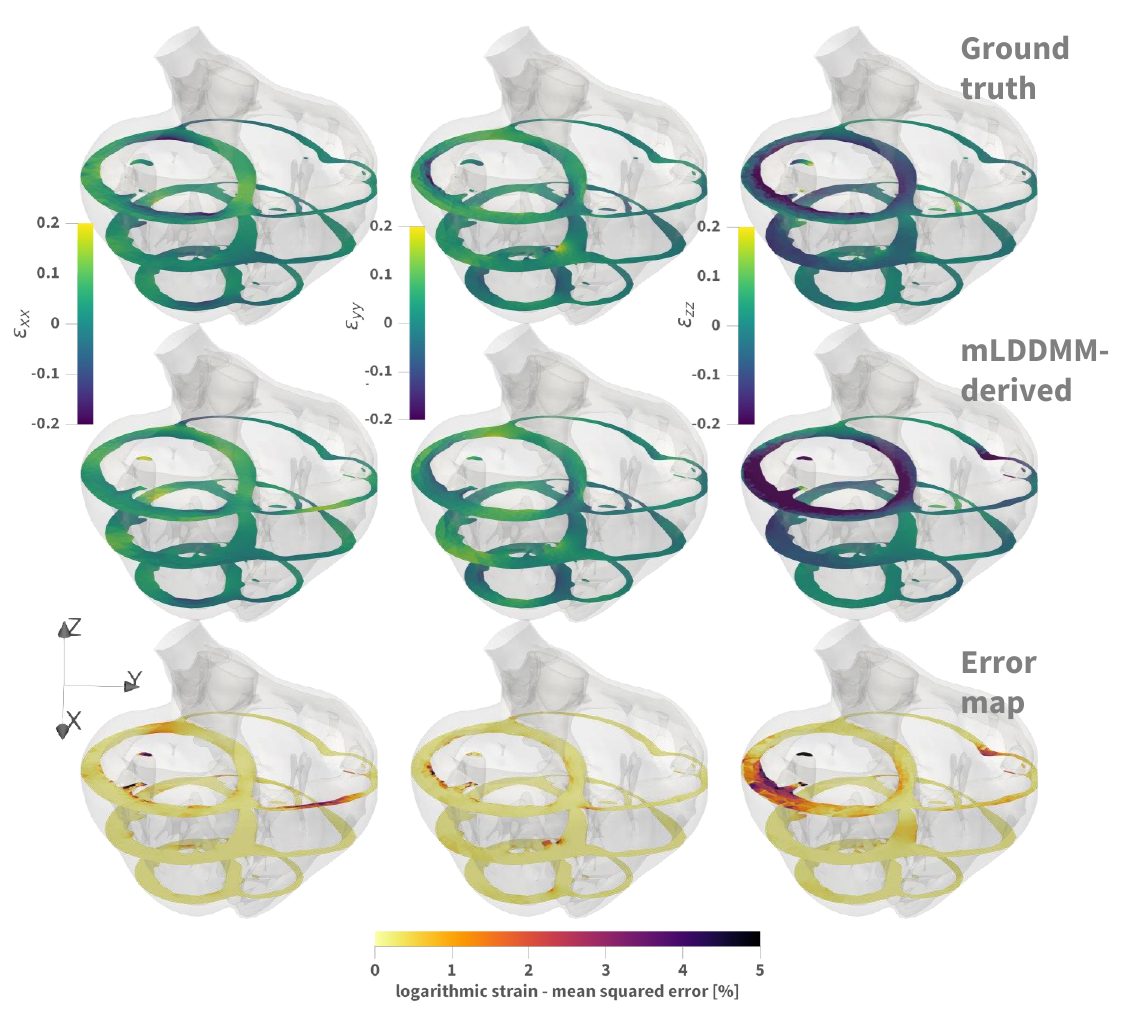}
    \caption{\textbf{mLDDMM strain estimation power, short-axis.} FE- and mLDDMM-derived logarithmic strain tensors components comparison, short-axis views.}
    \label{fig:le_comparison_slices_error}
\end{figure}

We analyze the strain components derived from FE and mLDDMM methods both in long axis and short axis planes, as shown in Figures \ref{fig:le_comparison_long_axis_errors} and \ref{fig:le_comparison_slices_error}.
In the long-axis plane, both the FE-computed and mLDDMM-derived x-directional strain maps show slightly positive strains in the mid-wall portion of the left free wall and negative strains in its basal portion, with mLDDMM estimates exhibiting slightly higher magnitudes. Similarly, we observe a strong agreement between the mLDDMM-derived y-directional spatiotemporal strain distribution and ground truth. The mLDDMM-derived z-directional strain aligns qualitatively with the ground truth, but some quantitative discrepancies in magnitude can be noticed, particularly near the papillary muscles. The MSE map confirms these trends, highlighting higher errors in the basal region of the left and right free walls. However, mean squared errors remain below 10\%. Notably, a large error occurs at the papillary muscle tip, likely due to its thin structure, which complicates mapping.

\smallskip
\noindent Similarly, in the short-axis plane, the FE-computed and mLDDMM-derived logarithmic strain tensors show strong agreement in the x- and y-directional strains. In the x-direction, both methods reveal slightly positive strains in the basal portion of the septum and left free wall and slightly negative strains in the apical region. The y-directional strain follows a similar trend, with consistent strain patterns along the septal and free walls across all slices. The respective MSE maps display error magnitudes ranging between 0 and 2\% in all planes, indicating very good agreement between the ground truth and the mLDDMM-based estimation. The mLDDMM-derived z-directional strain, however, exhibits some discrepancies with the ground truth, particularly in the basal slice. Despite this, both approaches reveal negative strains in the basal and mid-wall slices. The corresponding MSE map highlights these differences, though maximum errors in the basal slice remain below 5\%.

\subsection{Global MSE analysis metrics}
We complement our prior point-wise analysis with global metrics for the mean squared strain estimation error, specifically the 99th percentile values MSE, for each component. 
The results indicate highly accurate estimations for both the x- and y-directional strains, with the maximum error remaining below 5\%. 
Specifically, the 99th percentile MSE values for these directional strains amount to 4.641\% and 3.943\% respectively. 
In contrast, z-directional strain estimations show a higher error rate, reaching up to 6.864\%. 
This suggests greater challenges in accurately mapping strains in this direction, likely due to complex motion and deformation patterns that occur perpendicularly to the z-plane.
%
\section{Discussion}
Cardiac strain measurements serve as key biomarkers for early cardiac dysfunction diagnosis. As such, considerable effort has been dedicated to the development of Fourier-based and tracking-based methods for cardiac motion estimation \cite{TobonGomex2013,Genet2018,McLeod2012,Genet2023,Arratia-Lopez2023}. In this regard, recent studies have shown the importance of \textit{mechanically-}informed regularization of the deformation estimation problem for cardiac strain estimation \cite{Genet2018, McLeod2012, Arratia-Lopez2023}. In the domain of Fourier-based methods, WarpPINN has been recently proposed as a physics-informed neural network for heart deformation estimation, with near-incompressibility of the tissue enforced by penalizing the Jacobian of the deformation field \cite{Arratia-Lopez2023}. In the domain of tracking-based methods, a recently proposed regularization method based on the continuum finite strain formulation of the equilibrium gap principle was applied to image-based cardiac deformation estimation, ensuring minimal deviations from the mechanical equilibrium \cite{Genet2018}. Similarly, an extension of the log-domain demons algorithm was reported to allow the estimation of transformations that would obey physiological constraints of the cardiac tissue, including incompressibility and elasticity.
Likewise, our study introduces an innovative extension of the LDDMM framework towards tracking-based cardiac motion analysis and strain tensor extraction. Unlike previous approaches leveraging LDDMM for cardiac motion tracking, we include well-known mechanical constraints into the \textit{Deformetrica} framework for enhanced cardiac strain estimation.

\smallskip
\noindent\textbf{Framework enhancement.} 
The LDDMM framework has been previously employed for mesh-based motion tracking of anatomical structures \cite{Biffi2017, Risser2012}. Yet, such applications lacked volumetric displacement tracking of internal mesh nodes, critical for cardiac strain assessment. While recent efforts have been made to account for volumetric information retrieval from LDDMM-based frameworks \cite{Guan2024}, these works mainly focused on extending the framework towards growth tensor estimation and adjusted the LDDMM cost function for mesh quality purposes. In our mLDDMM approach, we regularize the cost function to take into account the natural nearly-incompressible behavior of myocardial tissue, ensuring an optimization approach that converges towards a unique mechanically sound estimated strain profile. 

\smallskip
\noindent\textbf{Volume displacement estimation.} 
We find a very good agreement between the FE- and mLDDMM-derived displacement fields. As shown in Figure \ref{fig:u_comparison}, the mLDDMM-estimated x- and y-, and z-directional displacements closely match the ground truth displacement fields, with minor residual discrepancies near the outflow tracts and in the z-direction. 
We conjecture these residual discrepancies to be due to the nature of the distal boundary conditions imposed on the proximal vasculature in the ground truth finite element simulation, which allows for some residual rigid body motion of the biventricular domain.

\smallskip
\noindent\textbf{Local strain tensor estimation.} 
Our method accurately estimates strain tensors on a complex cardiac contraction benchmark, as shown in Figures \ref{fig:le_comparison_long_axis_errors} and \ref{fig:le_comparison_slices_error}. 
Regional MSE analysis highlights localized errors in the basal portion of the left ventricle, with a greater mismatch in the z-directional strain. We report maximum localized error of 10\% in the long axis views, and of up to 5\% in the short axis slices. These error magnitudes are expected in these pipelines, as recently shown in a FE-based study on standardizing the validation of 4D cardiac strain imaging methods \cite{Mukherjee2024}. It is worth noting that we currently report both  FE-computed and mLDDMM-derived strains with respect to the Cartesian coordinate system. Transforming the strain tensor to the widely used radial-circumferential-longitudinal orientation basis will allow easier comparison with the existing literature \cite{Mukherjee2024, TobonGomex2013}. Finally, it is important to remark that while the magnitude of the strain field is modest, this might be attributed to the assumption of zero displacement at the reference mesh, which rules out the contribution of pre-stretch to the strain tensor in both the FE-derived ground truth and mLDDMM-based estimation.

\smallskip
\noindent\textbf{Limitations.} Current incompressibility penalty settings rely on empirical choices based on mesh characteristics. 
This is mainly due to the number of mesh nodes influencing the order of magnitude of the Varifold distance and the number of mesh elements influencing the order of magnitude of the incompressibility metric. 
Future refinements to our framework will explore improved normalization techniques for these loss terms as well as improving the standardization of the choice of these parameters given the mesh discretization at hand. 
Moreover, the current work does not consider the effect of noise on our proposed framework. In our future work, we aim to evaluate the effect of noise stemming from e.g., mesh distortion or misalignment artifacts on the performance of our method \cite{Berberoglu2019, Berberoglu2021}.

\smallskip
\noindent\textbf{Outlook} 
Future investigations will focus on comparing mLDDMM to existing benchmarks of clinical and mechanically regularized cardiac motion estimation \cite{TobonGomex2013}.
Furthermore, our framework could be extended to include additional microstructurally informed mechanical constraints in the spatial vector field, which guide the deformation within the ambient space and influence the proposed surface at each iteration \cite{Moscoloni2025}. 

\begin{credits}
\subsubsection{\ackname} This work was supported by the Research Foundation – Flanders, Fonds voor Wetenschappelijk Onderzoek –Vlaanderen (Grant No. 11PS524N, to B.M.) and European Union’s Horizon Europe Research and Innovation Program (VITAL - Grant No. 101136728, to P.S. and M.P.). 
\subsubsection{Disclosure of Interests} The authors have no competing interests that are relevant to the content of this article. 
\end{credits}
%
%
%
%
\printbibliography
\section*{Appendix}
\subsection*{A. mLDDMM estimation time}
All the mLDDMM computations were performed on a work station with an Intel i7-1370P with 32 GB RAM and an NVIDIA GeForce MX550. The estimation between the reference and the deformed configuration took about 1h. While the computation time is considerably higher than the computation time for \textit{regular} LDDMM, the framework can be further optimized to reduce the estimation time. Additionally, while  the high detail of the synthetic data served as a complex benchmark to showcase the robustness of the method in tensor extraction, the dimensionality of the discretization significantly influenced the computational time required for the estimation. Further work on diverse, less refined benchmarks should be conducted to evaluate the influence of the mesh complexity on computation time.
\end{document}